\documentclass[aps,prb,showpacs,superscriptaddress,twocolumn]{revtex4-1}
\usepackage{graphicx}
\usepackage{amsmath}
\usepackage{bm}
\usepackage{xcolor}

\begin{document}

\title{Evolution of individual quantum Hall edge states in the presence of disorder}

\author{Kai-Tong Wang}
\affiliation{School of Physics, Beijing Institute of Technology, Beijing 100081, China}

\author{Fuming Xu}
\affiliation{Shenzhen Key Laboratory of Advanced Thin Films and Applications, College of Physics and Energy, Shenzhen University, Shenzhen 518060, China}

\author{Yanxia Xing}
\email[]{xingyanxia@bit.edu.cn}
\affiliation{Beijing Key Laboratory of Nanophotonics and Ultrafine Optoelectronic Systems, School of Physics, Beijing Institute of Technology, Beijing 100081, China}

\author{Hong-Kang Zhao}
\email[]{zhaohonk@bit.edu.cn}
\affiliation{School of Physics, Beijing Institute of Technology, Beijing 100081, China}

\begin{abstract}

Employing the Bloch eigenmode matching approach, we numerically study the evolution of individual quantum Hall edge states with respect to disorder. As shown by the two-parameter renormalization group flow of the Hall and Thouless conductances, quantum Hall edge states with high Chern number $n$ are completely different from that of $n=1$ case. Two categories of individual edge modes are evaluated in a quantum Hall system with high Chern number. Edge states from the lowest Landau level have similar eigenfunctions which are well localized at the system edge and independent of the Fermi energy. On the other hand, at fixed Fermi energy, the edge state from higher Landau levels has larger expansion, which leads to less stable quantum Hall states at high Fermi energies. By presenting the local current density distribution, the influence of disorder on eigenmode-resolved edge states is vividly demonstrated.

\end{abstract}

   %Electronic transport phenomena in thin films

\maketitle
\section{INTRODUCTION}

Integer quantum Hall effect(IQHE) has attracted intensive research attention since proposed in 1980.\cite{Klitzing1980} Different from the mono-gapped system such as quantum anomalous Hall(QAH) and quantum spin Hall(QSH) systems, the band structure of IQHE system is multi-gapped, which is characterized by high Chern number $n$. As a result, the quantum phase transition of IQHE is of particular interest, such as the metal-insulator transition(MIT).\cite{Bao2015,Evers2008} To study the transition behavior of QH states, the two-parameter scaling theory was proposed\cite{Pruisken1988,Khmel1983} and numerical investigations were performed on a QH system to verify the two-parameter flow.\cite{Song2014,Werner2015} Besides, global phase diagram \cite{Kivelson1992} of IQHE predicted that there was no direct transition from general QH states to insulator under the influence of perturbations or disorders for high occupation case. But later experiments\cite{Song1997,Wang1994,Kravchenko1995} and theories\cite{Sheng1998,Sheng2000,Xie1996} indicated that direct MIT of $n>1$ QH state was achievable.

All the prominent features of MIT in QH systems are due to the appearance of multiple edge states at the device's boundaries\cite{Halperin1982,Buttiker1988}. These edge states originate from different Landau levels induced by external magnetic field, and the number of edge states equals the Chern number $n$. Unlike the QAH chiral edge states and QSH helical edge states\cite{Ren2016,Ren2017}, QH edge states are robust against any type of disorder, static or spin-dependent. During the metal-insulator transition of QH states, edge states from different Landau levels respond distinctly to disorder, since those gapped Landau levels are closely linked through edge states. Therefore, the influence of disorder on edge states from different Landau levels remains a fascinating issue, and the underlying mechanism triggers our research interest. For simplicity, the QH states discussed here and after all refer to IQHE.

In this work, we first investigate the two-parameter renormalization group(RG) flow of QH states in 2-dimensional lattice system based on a supercell system\cite{Niu1985,Hatsugai1999,Essin2007,Zhang2012}. The renormalization group flow for the $n=2$ Landau level is completely different from that of $n=1$, suggesting the fascinating MIT feature of QH systems with high Chern number. To study the phase transition of QH systems, we calculate the Hall and Thouless conductances\cite{Werner2015,Song2016,Nomura2008,Edwards1972} within the lattice gauge\cite{Werner2015}. Numerical results confirm that the $n>1$ QH state directly transforms into insulator but the critical transition disorder strength is smaller compared with the case of $n=1$. Furthermore, by adopting non-equilibrium Green's function(NEGF) formalism\cite{Wang2014} and Bloch eigenmode matching approach\cite{Khomyakov2005,Ando1991,Zhang2012a}, we calculate the eigenmode-resolved transmission coefficient of individual QH edge states with respect to disorder. The results show that, probability distractions of edge states from the lowest Landau level are always localized at the lattice edge despite of the variation of Fermi energy, which keeps these edge states robust and stable against disorder. On the other hand, the edge state originated from higher Landau levels has larger expansion across the system and they are more sensitive to disorder. The quality of QH states at the same high Chern number with different Fermi energy is also discussed. By showing the local current density distributions, the evolution of individual edge states with respect to disorder is intuitively displayed.

The rest of the paper is organized as follows. In Sec.\uppercase\expandafter{\romannumeral2}, the theoretical formalisms are introduced and we derive the transmission coefficients and the local current density for the propagating modes. Sec.\uppercase\expandafter{\romannumeral3} is the numerical results with discussions about our work. Finally, a brief summary is presented in Sec.\uppercase\expandafter{\romannumeral4}.

\section{MODEL AND FORMALISM}
In this section, the system Hamiltonian and related numerical formalisms are introduced.
\subsection{Hall and Thouless Conductances}
For a two-dimensional square lattice with external magnetic field, the Hamiltonian is expressed as\cite{Xing2008}
\begin{equation}
H = \sum_{\mathbf{i}} \epsilon_{\mathbf{i}} d_{\mathbf{i}}^{\dag}d_{\mathbf{i}}
- t \sum_{<\mathbf{i,j}>} e^{i \phi_{\mathbf{i}\mathbf{j}}} d_{\mathbf{i}}^{\dag} d_{\mathbf{j}}+ \text{H.c.} \label{eq1}
\end{equation}
where $d_{\mathbf{i}}^{\dag}$/$d_{\mathbf{i}}$ is the creation/annihilation operator for an electron on site $\mathbf{i}$, and $<\mathbf{i,j}>$ denotes nearest neighboring lattice sites. The random on-site potential $\epsilon_{\mathbf{i}}$ is uniformly distributed in the interval $\text{[-W/2,W/2]}$ where $\text{W}$ is the disorder strength, known as the Anderson-type disorder. $t$ is nearest-neighbor coupling strength, which is set as the unit of energy and disorder in the calculation. In the presence of a perpendicular magnetic field, an extra phase is induced in the adjacent coupling, which is defined as $\phi_{\mathbf{i}\mathbf{j}}=\frac{e}{\hbar}\int_{\mathbf{i}}^{\mathbf{j}}\mathbf{A}\cdot \mathbf{dl}$ with $\mathbf{A}$ the magnetic vector potential. In the numerical implementation, two different gauges are adopted. For the eigenmode-resolved transmission through a ribbon system, we choose the Coulomb gauge and the vector potential is simply $\mathbf{A}=[-By,0,0]$ in Cartesian coordinate. To calculate the Hall and Thouless conductances, we use the lattice gauge with special treatment on the lattice edge\cite{Werner2015}, which allows to reduce the system size at a fixed magnetic field. In the calculation, the extra phase through unit lattice measures the magnetic field strength. Spin degree of freedom is not considered in this work.

To calculate the transverse Hall conductance $g_H$, it is straightforward to adopt the relation between $g_H$ and the total Chern number
\begin{equation}
g_H=\frac{e^2}{h}\Sigma_n c_n
\end{equation}
where $c_n$ is the Chern number of the $n$th band and the summation is taken over all the bands below the Fermi energy. The Chern number defined in $k$ space can be written as\cite{Song2016,Fukui2007,Fukui2005,Fukui2007a}
\begin{equation}
c_n = \frac{1}{2\pi i}  \int_{BZ} d^2k \nabla \times \mathbf{A'} \nonumber
\end{equation}
with $\mathbf{A'}=<\psi_{nk}| \partial | \psi_{nk}>$ the Berry connection\cite{Thouless1982,Berry1984,Simon1983} and $\psi_{nk}$ the normalized wave function of the $n$th Bloch band.

On the other hand, the longitudinal conductance can be measured by the band sensitivity subject to changing in boundary conditions\cite{Braun1997,Edwards1972}. For a disordered system, Thouless and Edwards\cite{Braun1997,Edwards1972} proposed a relation between the average longitudinal conductance and the band curvature at the Fermi energy, which is called the Thouless conductance. At Fermi energy $E_f$, the Thouless conductance is expressed as\cite{Werner2015,Song2016}
\begin{equation}
g_{T}(E_f) = \frac{e^2}{h} < |\frac {\pi} {\Delta(E_f)} \frac{\partial ^2 E_f}{\partial k_{x}^{2}}| >
\end{equation}
where $\frac{\partial ^2 E_f}{\partial k_{x}^{2}}$ denotes the band curvature at $E_f$ and $\Delta(E_f)$ is the mean level spacing. The angle bracket stands for ensemble averaging over the disorder and $| ... |$ means taking the absolute value. Since both $g_H$ and $g_T$ are in unit of $\frac{e^2}{h}$, the unit is omitted in the numerical results shown below.

\subsection{Local Current Density and Eigenmode-resolved Transmission Coefficients}
For the lattice Hamiltonian $H$ shown in Eq.(\ref{eq1}), based on the Green's function formalism\cite{Jauho1994,Xing2010,Li2007} and the definition of current density\cite{Jiang2009,Xing2011}, the differential local current density vector at zero temperature reads
\begin{equation}
  d\mathbf{J}_{x/y}/d(eV) = \frac {1} {2}e[\hat{\rho}  \hat{v} + \hat{v} \hat{\rho}] \label{eq2}
\end{equation}
with
\begin{equation}
\begin{split}
&\hat{\rho} = \frac {1} {2\pi}  \mathbf{G}^{r} \mathbf{\Gamma}_{s} \mathbf{G}^{a} \\ \nonumber
&\hat{v} = -\frac {i} {\hbar}[\mathbf{r},\mathbf{H}] \nonumber
\end{split}
\end{equation}
where $\hat{\rho}$ is the nonequilibrium density matrix and $\hat{v}$ denotes the velocity matrix in the scattering region. $\mathbf{\Gamma}_{s} = i(\mathbf{\Sigma}_{s}^{r}-\mathbf{\Sigma}_{s}^{a})$ is the linewidth function and $\mathbf{\Sigma}_{s}^{r}$ denotes the retarded self energy of source lead $\bf{s}$, which can be generally calculated using the transfer matrix method\cite{Lee1981,Lee1981a}.

To distinguish different propagating modes, we employ the mode matching technique\cite{Khomyakov2005,Ando1991,Sanvito1999,Rungger2008}. For a system divided into slices with onsite Hamiltonian $H_{0}$ and hopping Hamiltonian $H_{1}$, the Bloch equation is written as
\begin{equation}
  (K_{0}+K_{1}e^{ik_{n}}+K_{-1}e^{-ik_{n}})\psi_{n} = 0 \\%\nonumber
\end{equation}
where $K_{a}=H_{a}-ES_{a}(a=0,\pm 1)$, $S$ is the overlap matrix and $\psi_{n}$ denotes the eigenvector of the $n-th$ eigenmode. Solving this equation, we can separate right-going and left-going modes. Selecting the propagating modes and substituting them into Eq.(\ref{eq2}), the linewidth function $\Gamma_s$ becomes
\begin{equation}
  \Gamma_{s} = \sum_{s_{n}} Q_{s_{n}} \frac {1}{v_{s_{n}}} Q_{s_{n}}^{\dagger} \label{eq3}
\end{equation}
with
\begin{equation}
\begin{split}
& Q_{s_{n}} = \mathbf{G_{00}^{r -1}} \psi_{s_{n}} \\ \nonumber
& v_{s_{n}} = i \frac {a}{\hbar} [\psi^{\dagger} K_{1} \psi e^{ika} - e^{-ika} \psi^{\dagger} K_{-1} \psi]  \nonumber
\end{split}
\end{equation}

Similarly, the eigenmode-resolved transmission coefficient can be written as
\begin{equation}
  T_{\alpha_m,\beta_n} = |t_{\alpha_m,\beta_n}|^2 \\%\nonumber
\end{equation}
where $t_{\alpha_m,\beta_n}$ is the transmission matrix element. Parameters $\alpha,\beta$ and $m,n$ label the leads and modes, respectively. The transmission matrix element $t_{\alpha_m,\beta_n}$ is calculated from\cite{Khomyakov2005,Ando1991}
\begin{equation}
t_{\alpha_{m},\beta_{n}}=\sqrt{|v_{\alpha_{m}}|} \tilde{\psi}_{\alpha_{m}}^{\dag} \mathbf{G}_{\alpha \beta}^{r}  Q_{\beta_{n}} \frac{1}{\sqrt{|v_{\beta_{n}}|}}
\end{equation}
Here $\mathbf{G}_{00}^{r}$ is the Green's function of an infinite ribbon. The eigenvector $\psi_m$ satisfies the orthonormalization condition $\tilde{\psi}_{n}^{\dag} \psi_{m} = \delta_{n,m}$. $v_{\alpha_m}$ is the group velocity of injecting or outgoing electrons of the $m-th$ eigenmode in lead $\alpha$. Detailed numerical procedures can be found in relevant references cited above.

\section{NUMERICAL RESULTS AND DISCUSSION}

\begin{figure}
\includegraphics[width=8.5cm, clip=]{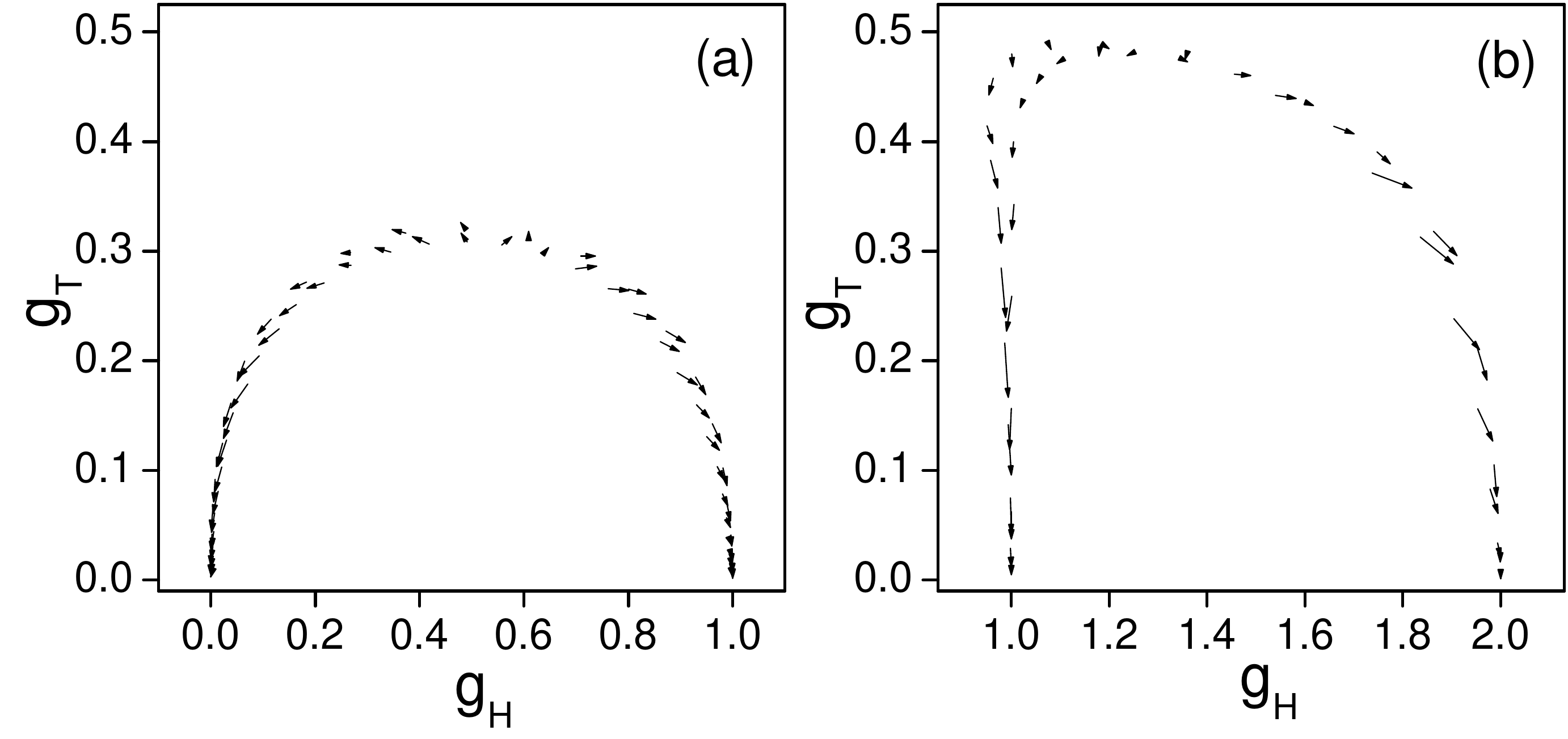}
\caption{ (Color online)The renormalization group flows extracted from finite size scaling, with system size ranging from $wid$=12 to 27 and disorder strength $W\in$ [1,1.5] for magnetic field $\phi=2\pi/9$. The arrows indicate the increasing in system size. The flows in panels (a) and (b) corresponds to Landau levels $n=1$ and $n=2$, respectively. The numerical results are averaged over $10,000$ random configurations.}
\end{figure}

To investigate the critical behavior in the topological phase transition, we calculate the two-parameter renormalization group(RG) flow of Hall conductance $g_H$ and Thouless conductance $g_T$ driven by the system size. In the presence of disorder, discrete Landau levels are broadened into Landau bands. A critical energy appears in the center of the Landau band, supporting an extended state and separating adjacent QH phases\cite{Werner2015}. Adopting the lattice gauge, the extra phase per unit lattice associated with the magnetic field is set as $\phi=2\pi/9$. The corresponding RG flows are shown in Fig.1. In Fig.1(a), we plot the RG flow related to the first Landau level, where the flow links the zeroth and the first QH states with $g_H = 0$ and $g_H = 1$, respectively. The arrows show the behaviors of $g_T$ and $g_H$ as the system size increases. Clearly, $g_T$ always decreases with the increasing of system size. On contrary, the Hall conductance decays to $g_H = 0$ at the left and grows to $g_H = 1$ at the right. As a result, a transition point corresponding to the critical energy emerges in the middle of the flow map, which is typical for mono-gapped topological insulators. The numerical results in Fig.1(a) are perfectly consistent with previous theoretical predictions\cite{Pruisken1988,Khmel1983}. On the other hand, the RG flow corresponding to the second Landau level ($n=2$) shown in Fig.1(b) is severely unsymmetric between $g_H = 1$ and $g_H = 2$. The critical transition point seriously deviates from the flow center but is close to the $g_H = 1$ QH state. The deviation and difference in Fig.1(b) indicate that the RG flow from the first Landau level is more robust against the disorder. Furthermore, these numerical results show that the quality of different Landau levels are not the same, which will have an impact on the evolution of QH states at metal-insulator transition(MIT) in the presence of disorder.

\begin{figure}
\includegraphics[width=8.5cm,clip=]{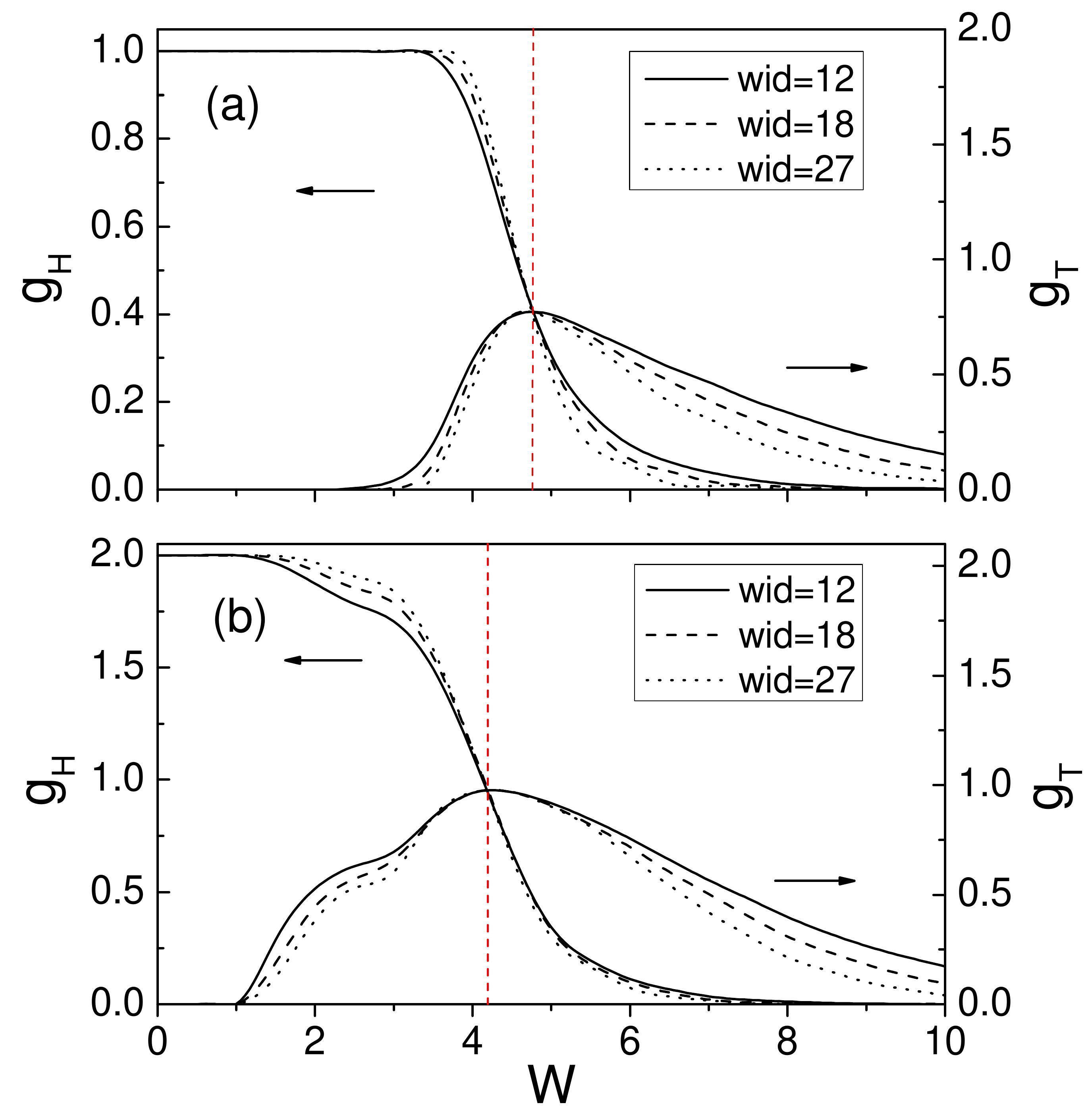}
\caption{ (Color online) Hall conductance $g_H$ and Thouless conductance $g_T$ versus the disorder strength $W$ for different system sizes. Panels (a) and (b) correspond to Fermi energies $E_f=-3$ and $E_f=-2$, which locate in the $n=1$ and $n=2$ QH plateaus, respectively. The red dashed lines indicate the critical transition disorder strengthes $W_1=4.7$ and $W_2=4.2$. The magnetic field is fixed at $\phi=2\pi/9$.}
\end{figure}

To explore the transition behaviors of QH states from conductor to insulator, we investigate the evolution of Hall and Thouless conductances with disorders at fixed magnetic field $\phi=2\pi/9$. The Fermi energy is first set as $E_{f}=-3$, which locates at the center of the $n=1$ QH plateau. The ensemble averaged Hall and Thouless conductances versus disorder strength $W$ are shown in Fig.2. At weak disorder, the Hall conductance is well quantized as $g_{H}=1$ and the Thouless conductance is zero. With the increasing of disorder strength, $g_H$ is swiftly reduced and eventually goes to zero, and $g_T$ quickly arises to maximum and then gradually decreases. For different systems sizes, all $g_H$ curves cross at a critical disorder point as indicated by the red dashed line in Fig.2(a). Meanwhile, maximal Thouless conductances appear at the same disorder, which are independent of system sizes. Except this critical point, the averaged $g_T$ depends on the system size and performs the behavior of localized states at large disorder\cite{Ando1983,Ando1984}. The critical disorder strength at $E_{f}=-3$ is $W_{1}=4.7$, which characterizes MIT of QH states. In Fig.2(b), we plot the dependence of $g_{H}$ and $g_{T}$ on the disorder strength $W$ at $E_{f}=-2$, which labels the $n=2$ QH states. Similar behaviors of $g_{H}$ and $g_{T}$ are observed, and we also find a critical disorder strength $W_{2}=4.2$ as shown by the red dashed line in Fig.2(b). The results prove that the $n=2$ QH states can directly transform to insulator without intermediate QH state. Besides, the critical transition disorder strength will decrease with the increase of Chern number, which means high QH states are more vulnerable to disorder. Together with the renormalized group flow shown in Fig.1, the properties of high QH states are more attractive since there are multi edge states originating from multi-gapped Landau levels. In the following, in order to demonstrate the evolution of individual edge states in the presence of disorder, we will intuitively visualize the edge states and exhibit the eigenmode-resolved transmission.

\begin{figure}
\includegraphics[width=8.8cm, clip=]{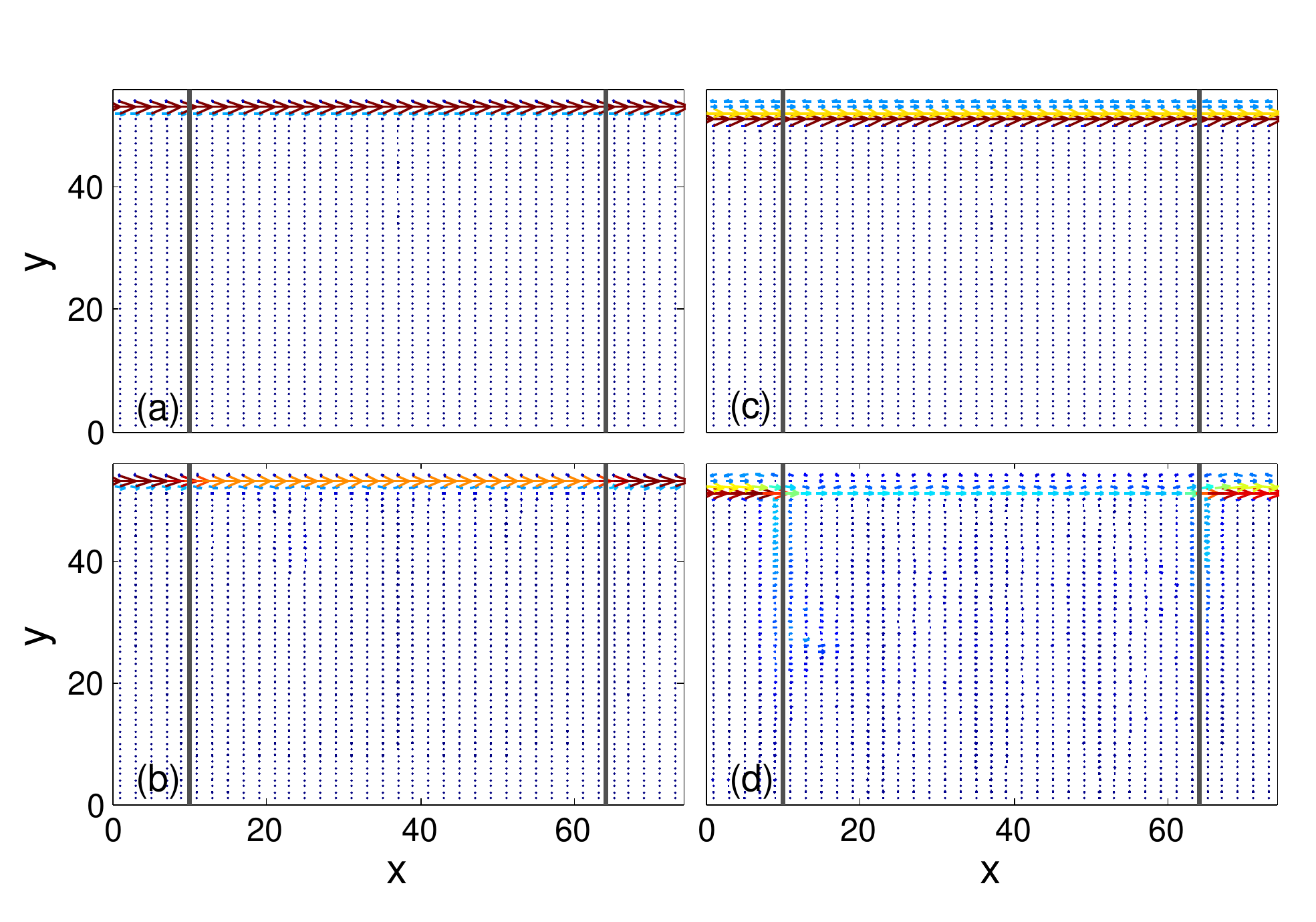}
\caption{ (Color online) Local differential current density distributions for both $n=1$ and $n=2$ QH states with fixed magnetic field $\phi=2\pi/9$. The arrow at the lattice site denotes the direction of the density vector and its length shows the density magnitude. The color of arrow changes from blue(light) to red(dark) means the increasing of magnitude. Other parameters in each panel: panel (a), $E_f=-3$ and $W=0$; panel (b), $E_f=-3$ and $W=2$; panel (c), $E_f=-2$ and $W=0$; panel (d), $E_f=-2$ and $W=2$. The width of the system is $wid=54$ lattices and $x$ is the transport direction. The results are averaged over $10,000$ configurations.}
\end{figure}

In real space, QH states display as a series of edge states in the system with finite size. The destruction of these edge states by disorder causes the metal-insulator transition. To intuitively visualize the edge states, we calculate the local current density for the $n=1$ and $n=2$ QH states and show the numerical results in Fig.3. The lattice system under investigation has width $wid=54$ and magnetic field $\phi=2\pi/9$. In Fig.3(a) and (c), we plot the local current density of QH states for energy $E_f=-3$($n=1$) and $E_f=-2$($n=2$) with disorder strength $W=0$. Clearly, the $n=1$ and $n=2$ QH states are perfectly localized at the upper edge of the lattice and no current density exists in the bulk, which is the signature of edge states. An obvious difference between the density patterns of two QH states is that the $n=2$ state has a larger expansion in $y$-direction, which is double-degenerate with $g_H=2$. When introducing disorder in the lattice, the local current density of both edge states are shown in Fig.3(b) and (d) with the same disorder strength $W=2$. The results are obtained through averaging over $10,000$ random configurations. It's obvious that both edge states are affected by the disorder. The edge-moving electrons injecting from the clean lead are scattered into the bulk of the central scattering region. The thick black lines indicate the border between clean leads and the disordered region. Comparing Fig.3(b) and (d), we find that the $n=1$ QH state still has a major edge portion with small density distributing in the whole scattering region. On the other hand, the edge state of $n=2$ is almost completely destroyed under the same disorder. We can confirm from these numerical results that, the edge state of $n=1$ QH state is more robust than the $n=2$ QH state. Hence a larger critical MIT disorder is expected for lower QH state. The $n=2$ QH state has double-degenerate edge states, which belongs to the $1st$ and $2nd$ Landau levels, respectively.

\begin{figure}
\includegraphics[width=8.cm, clip=]{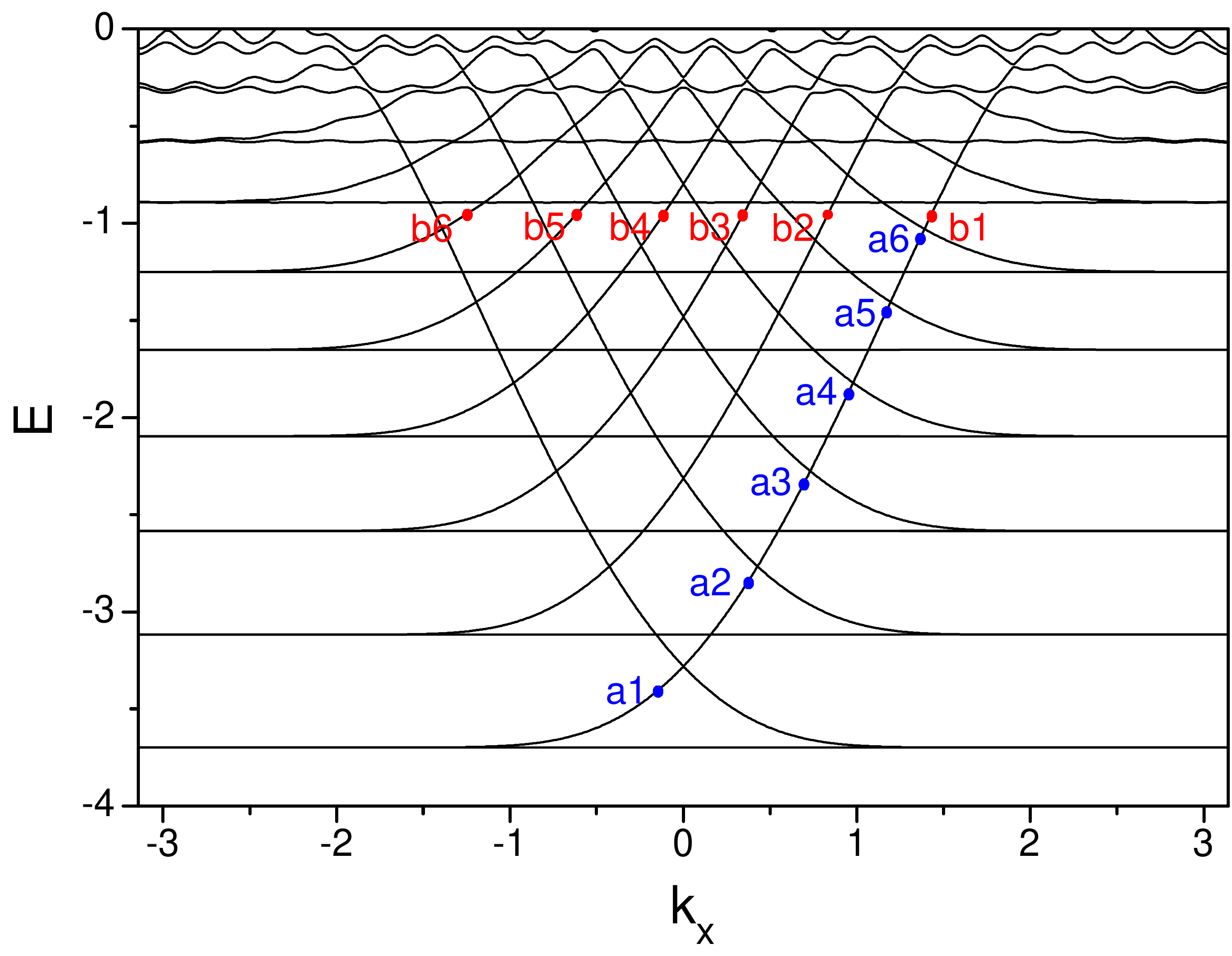}
\caption{ (Color online) The band structure of a lattice system in clean limit with magnetic field $\phi=2\pi/20$ and width $wid = 40$. The blue points(a1-a6) corresponds to the same Landau level in different gaps, and the red points(b1-b3) belongs to different Landau levels at the same Fermi energy $E_f=-0.95$.}
\end{figure}

To further understand the properties of edge states associated with different Landau levels, we separate degenerate edge states and investigate them individually for the high occupation case with Chern number $n>1$. In Fig.4, the band structure of a QH system is presented. The lattice under investigation has width $wid = 40$ with magnetic field $\phi=2\pi/20$. Due to the magnetic field, the original parabolic spectrum becomes highly degenerated and transforms to discrete Landau levels as shown in the figure. The edge states cross the gaps and link adjacent Landau levels. When scanning Fermi energy over the energy band, the edge state emerges one by one. These edge states fall into two distinct categories. At a high Fermi energy such as $E_f = -0.95$, there are six individual edge states, which are labeled as red points from b1 to b6. Among these edge states, b1 comes from the first Landau level, with b6 from the sixth. Since all the Landau levels are gapped and linked, these edge states may have different characteristics. On the other hand, the blue points (from a1 to a6) are chosen from the same Landau level(the lowest) but locate in the center of different gaps with various injecting energies. The edge qualities of these two categories will be intensively studied in the following. Notice that these energy points are non-degenerate.

\begin{figure}
\includegraphics[width=8.5cm, clip=]{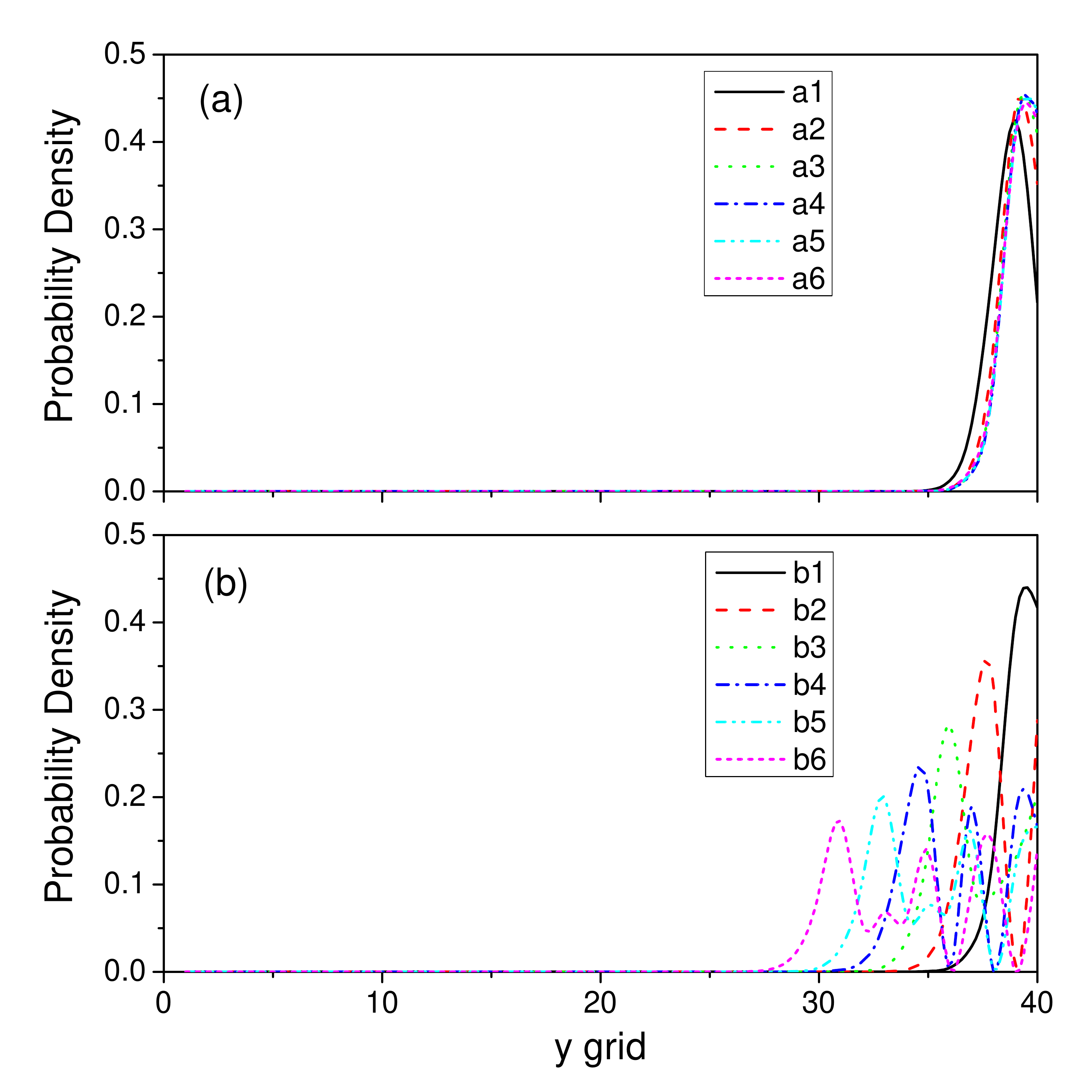}
\caption{ (Color online) The probability density distributions of edge states corresponding to the crossed points in Fig.4. Panel (a): density distributions of edge states from the lowest Landau level, and the Fermi energies lie in the center of each gap. Panel (b): density distributions of edge states from different Landau levels with the same fermi energy $E_{f}=-0.95$. The lattice width is $wid = 40$.}
\end{figure}

In Fig.5, we present the lattice probability densities of two series of edge states, the blue points a1-a6 and the red points b1-b6 in Fig.4. The probability densities of the blue points, i.e., edge states from the same Landau level with different energies are plotted in Fig.5(a). The result shows that in the absence of disorder, edge states from the lowest Landau level for different Fermi energies present the same behavior and have almost identical distributions. The eigenfunctions are strictly restrained on the boundary, acting as edge modes. On contrary, in Fig.5(b), the probability densities of edge states from different Landau levels with the same energy $E_f=-0.95$ have distinct distributions. Clearly, density distribution of the eigenmode from the lowest Landau level is perfectly localized at the lattice edge as indicated by the black solid line. Starting from b2, node appears in the eigenfunction of edge states and local minimums grow in the probability distributions in Fig.5(b). With the increasing of Chern number $n$, the probability distribution gradually spread into the lattice center, which implies that the quality of corresponding edge state decays. For instance, the density distribution corresponding to b6 edge state expands to one fourth of the lattice in $y$-direction. These numerical facts reveal the generic difference between edge states from the two categories. In the following, we will show the properties of these edge states in the presence of disorder, accompanies with local differential current density distributions defined in Eq.(\ref{eq2}).

\begin{figure}
\includegraphics[width=8.5cm, clip=]{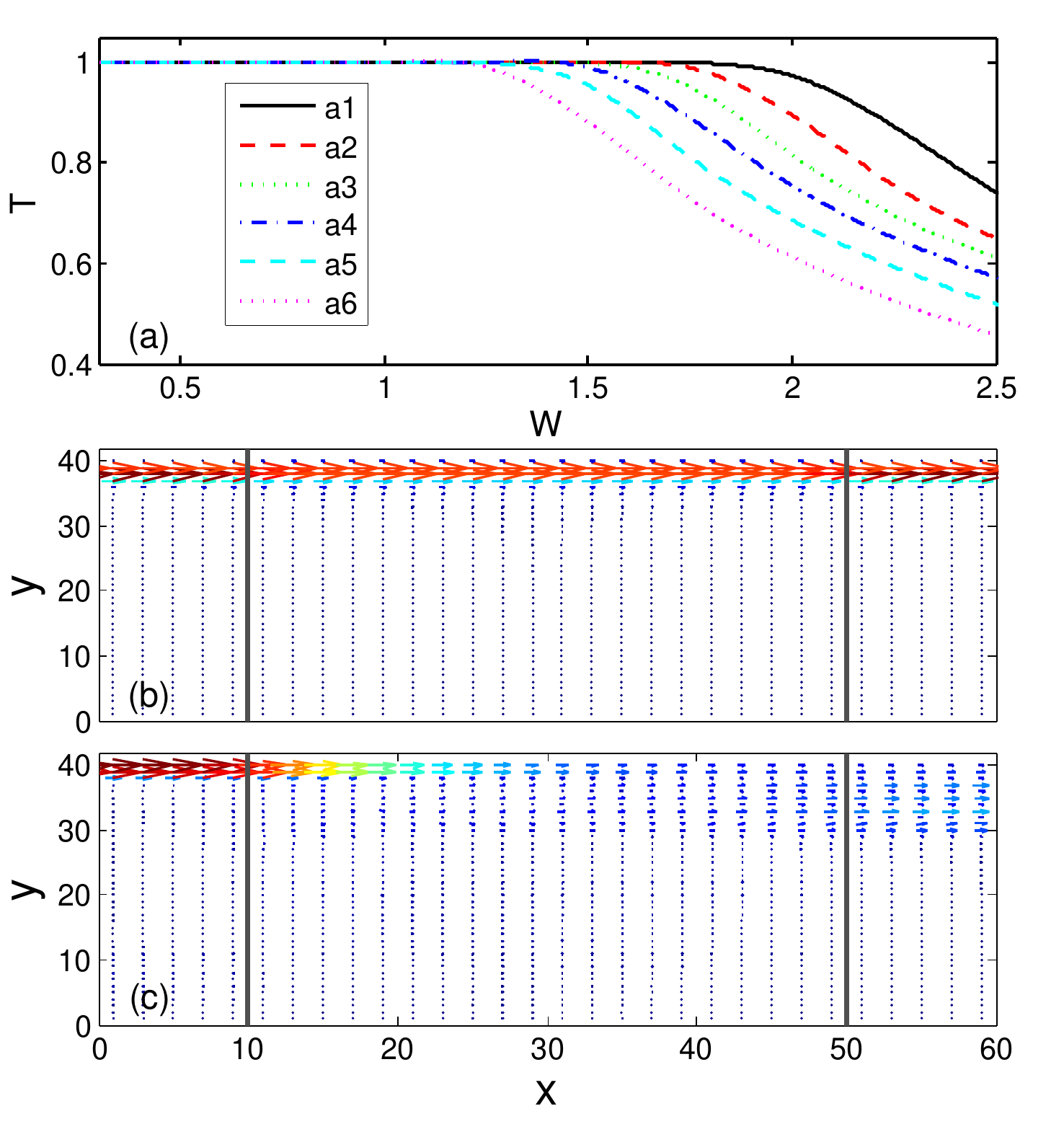}
\caption{ (Color online) Panel (a) plots the eigenmode-resolved transmission coefficients versus disorder for the first Landau level, with Fermi energies locating at the center of different gaps, as shown by blue points (a1-a6) in Fig.4; Panels (b) and (c) are the local differential current density distributions corresponding to edge states in the $1st$ and $6th$ gaps with disorder strength $W=1.5$. The magnetic field is $\phi=2\pi/20$ and the lattice has width $wid=40$, which are same as in Fig.4. }
\end{figure}

With the mode matching method, we are able to study the properties of individual edge states against disorder. Specifically, the eigenmode-resolved transmission coefficient $T$ of each propagating mode as well as its current density distribution in the lattice can be numerically calculated. We first study the evolution of individual edge modes from the first Landau level with different energies. We focus on the lowest 6 gaps as shown by blue points in Fig.4. Here, the injecting energy is set at the center of each gap. The eigenmode-resolved transmission coefficients for different energies versus disorder strength $W$ are shown in Fig.6(a). As presented in Fig.5(a), the eigenfunctions of edge states from the same Landau level are almost identical. However, from Fig.6(a), We can see that these edge states respond differently to large disorder. The eigenmode-resolved T satisfies the order $T_{a1} > T_{a2} > T_{a3} > T_{a4} > T_{a5} > T_{a6}$ in the whole disorder range. We also plot the local differential current density distributions with different energies in Fig.6(b) and (c). It is obvious that at the same disorder $W=1.5$, the edge state with lowest energy survives well, and the $6th$ edge mode has been destructed in the central scattering region. Since the energy gap between adjacent Landau levels becomes more and more narrow at high occupation, the edge state from the same Landau level with higher energy is more vulnerable to disorder.

\begin{figure}
\includegraphics[width=8.5cm, clip=]{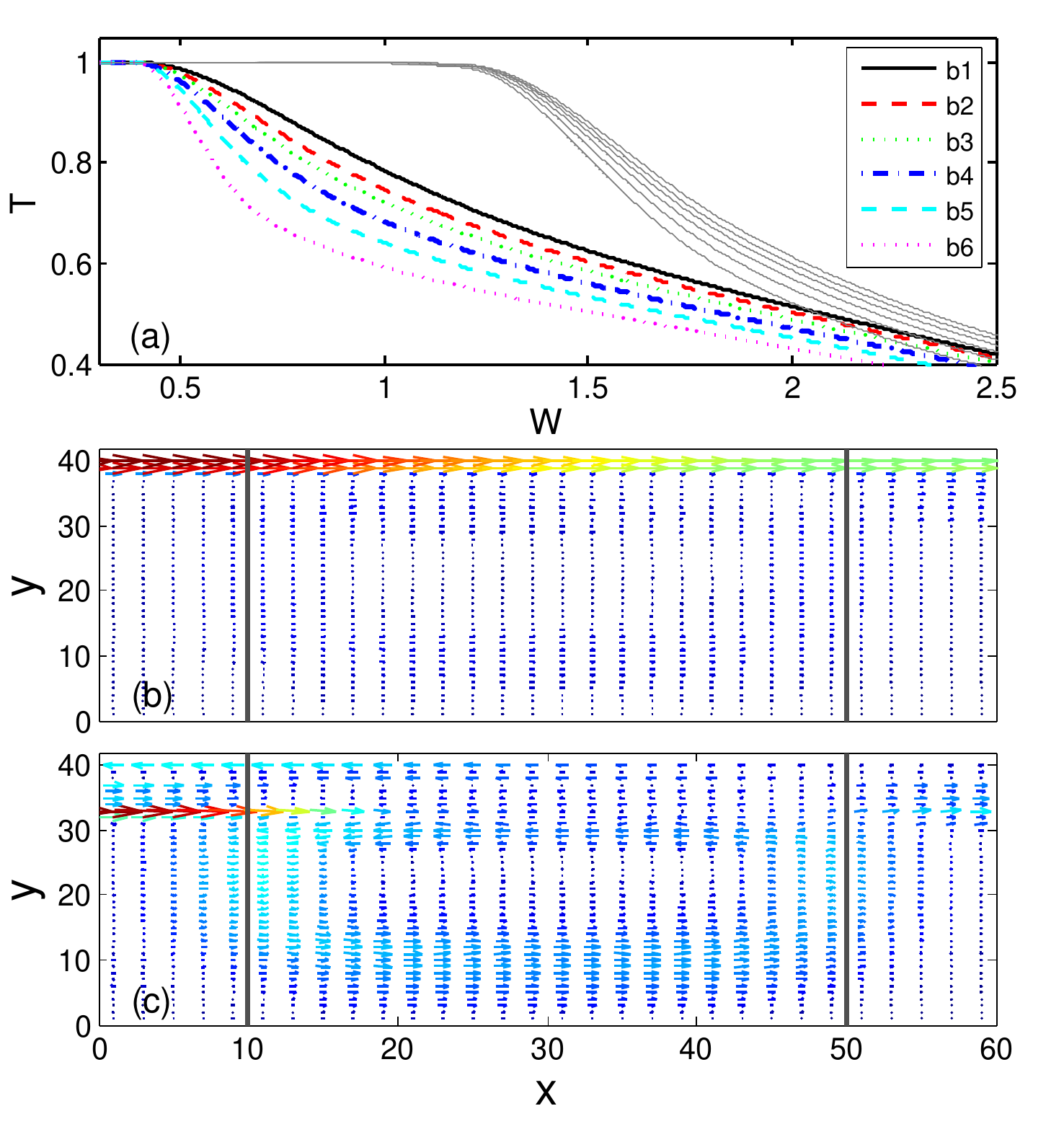}
\caption{ (Color online) Panel (a) shows the eigenmode-resolved transmission coefficients as a function of the disorder strength $W$ for fixed Fermi energies $E_{f1}=-1.07$ (grey lines) and $E_{f2}=-0.95$ (color lines), respectively. $T_{b1} > T_{b2} > T_{b3} > T_{b4} > T_{b5} > T_{b6}$ is observed in the whole disorder window for both Fermi energies. Panels (c) and (d) are the local differential current density distributions of eigenmodes corresponding to the $1st$ and $5th$ Landau levels of the $n=6$ QH state with $E_f=-0.95$ and disorder strength $W=0.7$. Other parameters are the same as in Fig.6. }
\end{figure}

Finally, we fix the Fermi energy and investigate edge modes from the same QH state. The $n=6$ QH states with two Fermi energies are evaluated and the eigenmode-resolved transmission coefficient $T$ versus disorder strength $W$ are presented in Fig.7(a). The data are obtained through averaging over $10,000$ configurations. Here we have chosen two Fermi energies, $E_{f1}=-1.07$ (gray lines) and $E_{f2}=-0.95$ (color lines), which are at the the center and top edge of the $6th$ energy gap, respectively. For weak disorder, all transmission coefficients are well quantized as $T=1$. With the increasing of disorder, the eigenmode-resolved $T$ starts to decrease gradually. A remarkable contrast is that, the resolved transmissions at $E_{f1}$ is always larger that that at $E_{f2}$ for the same disorder, which suggests that with the same high Chern number the QH state at the gap center is more robust than that close to the top edge of the same gap. Another intuitive fact is, for both $E_{f1}$ and $E_{f2}$, the order $T_{b1} > T_{b2} > T_{b3} > T_{b4} > T_{b5} > T_{b6}$ at the same $W$ is observed in the entire disorder window. This fact shows that, for the same Fermi energy, the resolved edge state of a lower Landau level is more stable than that of a higher Landau level. This result is consistent with the conclusion in Fig.5(b), since the edge state from a higher Landau level has larger expansion and is more easily affected by disorder. It is noticeable that, the resolved $T$ for the QH state at the top gap edge(color lines) is more easily to be distinguished from each than those for the QH state at gap center(gray lines), suggesting that MIT of this sensitive QH state is mainly caused by the edge states from high Landau levels. In Fig.7(b) and (c), we show the local current density distributions of edge states b1 and b5 for $E_{f2}=-0.95$ of the $n=6$ QH state. In the clean lead ($x \in [0,10]$), the $1st$ edge state is sharply localized at the lattice edge and the $5th$ one has a larger expansion in $y$-direction. In the scattering region with the same disorder $W=0.7$, the $1st$ edge mode is well preserved while the $5th$ edge mode is destroyed and spread in the whole region. These evidences clearly show for the QH state with high Chern number the edge state from higher Landau level is less stable.
\\

\section{CONCLUSION}

In summary, we have studied the evolution of edge states in disordered 2-dimensional quantum Hall systems. The two-parameter renormalization group flow shows that the QH state with high Chern number $n>1$ is completely different from that of $n=1$. By adopting non-equilibrium Green's function formalism and Bloch eigenmode matching approach, evolution of eigenmode-resolved edge states with respect to disorder is studied in detail. We have verified that for the lowest Landau level, edge states injecting from different band gaps have similar probability distributions. For a quantum Hall state at fixed Fermi energy with multiple edge states, the edge state from a lower Landau level is more localized at the system edge, and always more robust against disorder. On contrary, edge modes from higher Landau levels exhibit larger expansion across the system, and more sensitive to disorder. Furthermore, the quality of QH states with the same high Chern number heavily depend on the Fermi energy. When $E_f$ is at the top gap edge, the eigenmode-resolved transmission is more easily to be distinguished from each other than the gap center case. By presenting the local current density distributions, the evolution of eigenmode-resolved edge states in the presence of disorder is intuitively visualized.

$${\bf ACKNOWLEDGMENTS}$$
This work is financially supported by the National Natural Science Foundation of China (Grants No. 11674024 and No. 11504240).

\bibliographystyle{apsrev}

%---------------------------------------------------------------------------

\end{document}